# Electric-field-controlled directional growth of ferroelectric domains in multiferroic BiFeO$_3$ films

T. H. Kim (김태헌),[1] S.-H. Baek,[2] S. M. Yang,[1] S. Y. Jang,[1] D. Ortiz,[2] T. K. Song,[3] J.-S. Chung,[4] C.-B. Eom,[2] T. W. Noh,[1] and J.-G. Yoon[5,a]

[1]*ReCOE & FPRD, Department of Physics and Astronomy, Seoul National University, Seoul 151-747, Korea*

[2]*Department of Materials Science and Engineering, University of Wisconsin, Madison, WI 53706, USA*

[3]*School of Nano and Advanced Materials Engineering, Changwon National University, Changwon, Gyengnam 641-773, Korea*

[4]*Department of Physics and CAMDRC, Soongsil University, Seoul 156-743, Korea*

[5]*Department of Physics, University of Suwon, Hwaseong, Gyeonggi-do 445-743, Korea*

---

[a] Electronic mail: jgyoon@suwon.ac.kr

We describe the directional growth of ferroelectric domains in a multiferroic BiFeO$_3$ thin film, which was grown epitaxially on a vicinal (001) SrTiO$_3$ substrate. A detailed structural analysis of the film shows that a strain gradient, which can create a symmetry breaking in a ferroelectric double well potential, causes ferroelectric domains to grow with preferred directionality under the influence of an electric field. Our results suggest the possibility of controlling the direction of domain growth with an electric field by imposing constraints on ferroelectric films, such as a strain gradient.

The control of domain walls (DWs) in ferroic materials is currently an important issue in the potential application of DW motion to microelectronic devices.[1-4] It has recently been suggested that controlled movement of DWs in magnetic nanowires can be utilized for nonvolatile high-performance racetrack memory.[2] Additionally, DW-induced conductivity modulation has been observed in an insulating ferroelectric (FE) thin film, providing a basis for new device concepts using FE DWs.[4] Such DW devices require precise control of both the motion and location of the DWs.

In magnetic nanowires, the direction of the spin-polarized current controls the direction of DW motion.[2,5] When a spin-polarized current is passed through a DW, DW motion results from a spin-torque applied to the spin angular moments in the DW, which were transferred by the current. On the other hand, no reliable technique has been discovered to control the direction of FE DW motion.

In this paper, we describe the directional growth of FE domains in multiferroic $BiFeO_3$ (BFO) capacitors, which were grown epitaxially on vicinal (001) $SrTiO_3$ (STO) substrates. Unlike most other FE films, which always show isotropic DW growth, our BFO films exhibited FE domain growth in a particular direction under a given electrical bias. In addition, reversing the direction of the electrical bias reversed the direction of DW growth. We explain these surprisingly symmetry-broken DW growth behaviors in terms of a strain gradient, which can cause an asymmetry in a local FE double well potential. Our results indicate the possibility of controlling the direction of DW motion.

We grew epitaxial BFO(200 nm)/SrRuO$_3$(100 nm) heterostructures on vicinal (001) STO substrates with a 2° miscut toward [100] by using off-axis rf magnetron sputtering.[6,7] The SrRuO$_3$ layer was used as a bottom electrode. We formed 40 nm-thick Pt top electrodes (100 $\mu$m in diameter) by rf sputtering and photolithography. To monitor the movement of FE DWs under the influence of a uniform external electric field in the time scale between 1 and 200 $\mu$s, we used modified piezoresponse force microscopy (PFM). Details of this technique are discussed elsewhere.[8,9] To observe DW movement in a reasonable time scale, we subjected the BFO capacitors to an external bias of the same magnitude as the coercive voltage. The out-of-plane piezoelectric responses were recorded under an *ac* bias of 0.3 V$_{rms}$ at 19.1 kHz.

Figure 1(a) shows the time-dependent evolution of FE domains at the same site under a negative electric bias applied to the top electrode. The sequential PFM images show that the domain growth is highly anisotropic, contrary to most earlier results on FE thin films, which exhibited isotropic DW growth.[8,10-12] To display this directional growth more clearly, overlaid images of domains taken at different time intervals are shown in Fig. 1(c). When we applied a positive electric bias, we observed that the direction of DW movement was reversed, as shown in Figs. 1(b) and 1(d). These results clearly imply that the direction of DW movement can be controlled by the polarity of the electric field.

We found that the vicinality of the substrate should be a key factor in controlling the direction of DW motion. The direction of domain growth is along the [100] ([-100]) direction of the STO substrate under a negative (positive) bias field. Using x-ray reciprocal space mappings (RSMs), the [100] and [010] directions were determined to be identical to the *a*- and *b*-axes of a BFO unit cell, respectively. Also, the [100] direction is the same as the downhill miscut direction of the substrate (whose x-ray diffraction results are not shown here), indicating the importance of substrate vicinality. Note that bulk BFO has a rhombohedral crystal structure. On non-vicinal STO(001) substrates, BFO films should experience a biaxial in-plane strain, which would lead to identical *a*- and *b*-axis lattice constants.[7] On the other hand, on vicinal STO substrates, our BFO films were found to have different lattice constants along the *a*- and *b*-axes, indicating the existence of additional in-plane strains.[7] We argue that our vicinal STO substrate should provide a directional stress to the rhombohedral BFO unit cells, especially near the step edges. [We also performed PFM studies on BFO films grown on vicinal STO(111) substrates, which do not have a monoclinic distortion caused by underlying substrate strain (results not shown here). In this case, the domain growth was found to be nearly isotropic.]

In order to confirm the existence of directional stress and the resulting strains, we measured the depth profiles of the in-plane lattice constants in miscut BFO(001) films. High-resolution grazing incidence in-plane x-ray diffraction (GID) experiments were carried out at the 10C1 beamline of the Pohang Accelerator Laboratory in Pohang, Korea, using a 6-circle diffractometer. We varied the grazing incident angles from 0.2° to 1.0°. As the angle of the incident x-ray increases, its penetration depth $\lambda$ also increases.[13] Figure 2(a) displays the changes in the lattice constant along the *a*-axis (toward the miscut direction) and *b*-axis (perpendicular to the miscut direction) in terms of $\lambda$. The *a*-axis lattice constant shows little change down to ~80 nm from the surface of the film. Its value is consistent with the bulk value for BFO, indicating that the BFO film is fully relaxed along the *a*-axis. On the other hand, the *b*-axis lattice constant abruptly decreases when $\lambda$ is larger than ~50 nm. In other words, along the *b*-axis, the BFO film is partially strained by the substrate near the interface, but it becomes fully relaxed near the film surface. [On the other hand, the anisotropic in-plane strain states do not exist in BFO(001) films grown on non-vicinal STO substrates.[7]]

The strongly strained state along the *b*-axis may come from the step edges of the vicinal substrate. It is highly likely that the unit cells near the step edges are more easily strained than those far from the step edges. This could lead to an in-plane strain gradient on the terraces. Our GID data also indicate overall lattice relaxation with increasing film thickness, and hence, our BFO(001) film should have strain gradients in the in-plane as well as the out-of-plane directions, as is schematically depicted in Fig. 3(a).

The observed strain gradients can create non-uniform local internal fields via the piezoelectric effect.[14,15] As Fig. 3(b) indicates, the strain states of BFO unit cells would change gradually from strongly strained states to less strained or fully relaxed states as the position of the cells moves farther from the step edges. With a larger strain state, the induced local internal field becomes larger and the double well potential for FE polarization becomes more tilted.[16] Fig. 3(c) shows the variation of the double well potential in terms of the position. Under positive (negative) bias, the energy barrier height for polarization switching becomes smaller (larger) in the [100] direction, and thus DWs should move along the [-100] ([100]) direction. Consequently, the existence of strain gradients in our BFO(001) films on vicinal STO substrates enables us to control the direction of DW movement via the polarity of the external bias.

The strain gradients inside a BFO film can also significantly affect FE domain dynamics, such as nucleation sites and switching time. Figures 1(a) and 1(b) show that the initial nucleation sites are quite different for the negative and positive biases. We found that the number of initial nucleation sites for a positive bias was about ten times larger than that for a negative bias. Note that most of the nucleation in FE thin films should occur through an inhomogeneous process, possibly due to structural defects.[9] For DW motion along [-100] under a positive bias, the relaxed region should provide the nucleation sites. Because the relaxed region can have more structural defects, more nucleation sites should exist for a positive bias.

Figure 2(b) shows the area of switched polarization in the PFM images of Figs. 1(a) and (b). Under the negative bias, ~50 $\mu$s were required for half of the polarization to become switched. On the other hand, ~200 $\mu$s were required under the positive bias, in spite of the much higher nucleation density. This indicates that DWs can propagate about four times faster under a negative bias than under a positive bias. As Fig. 3(c) indicates, the energy barrier height of the strained region under the negative bias was smaller than those of the relaxed region under the positive bias. Accordingly, DW propagation should be faster under a negative bias.

In summary, we can control the direction of domain growth in BiFeO$_3$ films grown on vicinal SrTiO$_3$(001) substrates by changing the polarity of an applied bias. We attribute the directional domain growth to a non-uniform local internal field generated by anisotropic strain gradients developed inside the $BiFeO_3$ films. We believe that our approach, based on the substrate's vicinality, can be extended to other ferroelectric materials with bulk rhombohedral symmetry.

This research was supported by the Basic Science Research Program through the National Research Foundation of Korea (NRF) funded by the Ministry of Education, Science and Technology (No. 2009-0080567). The experiments at the Pohang Light Source were supported by MEST and POSTECH. C.B.E. gratefully acknowledges the financial support of the Office of Naval Research through grant N00014-07-1-0215, the National Science Foundation through grant ECCS-0708759, and a David & Lucile Packard Fellowship.

FIG. 1. (Color online) Out-of-plane PFM phase images of successive FE domain evolution under external (a) negative (-4.7 V) and (b) positive (3.3 V) biases after a poling process to make the capacitor a single domain. All of the PFM phase images were obtained in the same region. The scan size is 30×30 $\mu m^2$. (c) and (d) were obtained by superposing PFM phase images from (a) and (b). FE domains grow in a particular direction, indicated by the open arrows, depending on the polarity of the applied field. The solid dots indicate the regions where DWs become strongly pinned. ⊙ and ⊗ represent the direction of the negative and positive biases applied to the film, respectively.

FIG. 2. (Color online) (a) Depth profiles of lattice constants along the *a*-axis (toward the miscut direction) and the *b*-axis (perpendicular to the miscut direction) in the miscut BFO(001) films obtained from GID measurements and RSMs. The solid line and the dashed line indicate the lattice constants of bulk STO and BFO, respectively. (b) Time-dependent switched area, calculated from the PFM phase images displayed in Figs. 1(a) (negative bias) and 1(b) (positive bias).

FIG. 3. (Color online) (a) A schematic diagram of the anisotropic strain gradient induced

by the substrate's vicinality. The magenta and white colors correspond to the strongly strained and relaxed states, respectively. (b) A top view of the strain variation on a terrace. (c) Variation of the double well potential due to the strain-induced local internal field. Energy barriers for polarization switching in strongly strained (magenta), weakly strained (light magenta), and relaxed (white) BFO unit cells is schematically illustrated. The solid (black), dashed (red), and dashed-dot (blue) curves represent the double well potentials for the polarization states under zero, positive, and negative external bias fields, respectively.

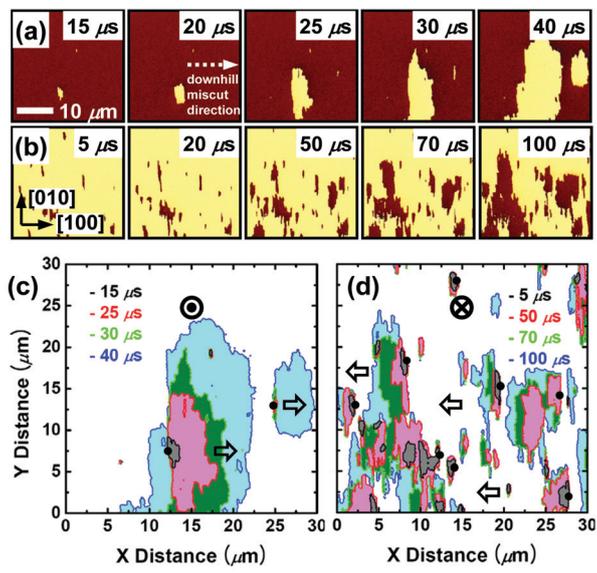

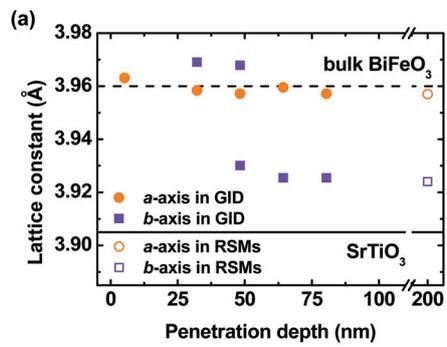
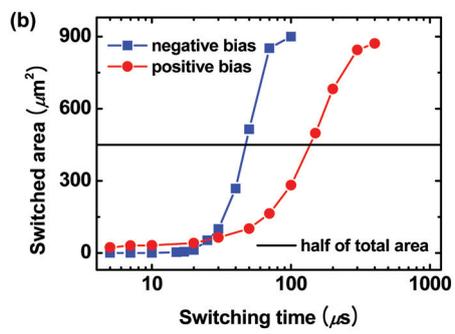

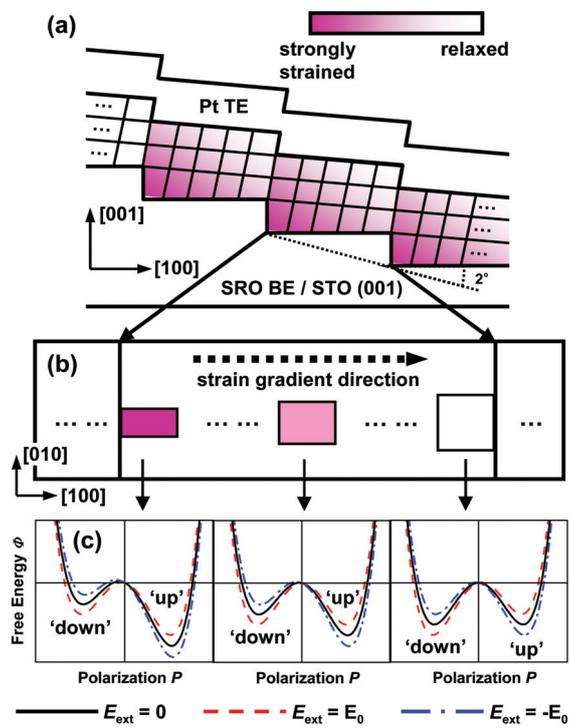